\documentclass{article}

\PassOptionsToPackage{numbers, compress}{natbib}

\usepackage[final]{tackling_climate_workshop_style}

\usepackage[utf8]{inputenc} 
\usepackage[T1]{fontenc}    
\usepackage{url}            
\usepackage{booktabs}       
\usepackage{amsfonts, amsmath}       
\usepackage{nicefrac}       
\usepackage{microtype}      

\usepackage{xcolor}
\usepackage[pdftex]{graphicx}
\usepackage{subcaption}
\usepackage{wrapfig}

\title{Surrogate Neural Networks to Estimate Parametric Sensitivity of Ocean Models}

\author{%
  Yixuan Sun\\ 
  Mathematics and Computer Science\\
  Argonne National Laboratory\\
  Lemont, IL 60439 \\
  \texttt{yixuan.sun@anl.gov} \\
  \And
  Elizabeth Cucuzzella \\
  Tufts University \\
  Medford, MA 02155\\
  \texttt{Elizabeth.Cucuzzella@tufts.edu} \\
  \AND
  Steven Brus \\
  Mathematics and Computer Science\\
  Argonne National Laboratory\\
  Lemont, IL 60439 \\
  \texttt{sbrus@anl.gov} \\
  \And
  Sri Hari Krishna Narayanan \\
  Mathematics and Computer Science\\
  Argonne National Laboratory\\
  Lemont, IL 60439 \\
  \texttt{snarayan@anl.gov} \\
  \And
  Balu Nadiga \\
  Los Alamos National Laboratory\\
  Los Alamos, NM 87545 \\
  \texttt{balu@lanl.gov} \\
  \And
  Luke Van Roekel \\
  Los Alamos National Laboratory\\
  Los Alamos, NM 87545 \\
  \texttt{lvanroekel@lanl.gov} \\
  \And
  Jan H\"uckelheim \\
  Mathematics and Computer Science\\
  Argonne National Laboratory\\
  Lemont, IL 60439 \\
  \texttt{jhuckelheim@anl.gov} \\
  \And
  Sandeep Madireddy \\
  Mathematics and Computer Science\\
  Argonne National Laboratory\\
  Lemont, IL 60439 \\
  \texttt{smadireddy@anl.gov} \\
}

\begin{document}

\maketitle

\begin{abstract}
  Modeling is crucial to understanding the effect of greenhouse gases, warming, and ice sheet melting on the ocean. At the same time, ocean processes affect phenomena such as hurricanes and droughts. Parameters in the models that cannot be physically measured have a significant effect on the model output. For an idealized ocean model, we generated perturbed parameter ensemble data and trained surrogate neural network models. The neural surrogates accurately predicted the one-step forward dynamics, of which we then computed the parametric sensitivity.
\end{abstract}

\section{Introduction}
 The oceans act as an important brake on anthropogenic climate change by absorbing carbon dioxide and atmospheric heat. At the same time, ocean processes play an important role in phenomena such as hurricanes and droughts.
Much effort is devoted to modeling and understanding the behavior of the ocean under various scenarios~\cite{deyoung2004challenges, semtner1995modeling, yan2018underestimated}. Of particular interest are the long-term changes in critical ocean circulation patterns such as the Atlantic Meridional Overturning Circulation (AMOC), which could have wide-ranging climate impacts. The AMOC is responsible for the northward heat transport throughout the entire Atlantic Ocean and is therefore an important process to accurately represent in Earth system models. Understanding the stability of this circulation is critical to our ability to predict the conditions that could cause a collapse in AMOC.
 
We are interested in understanding the sensitivities of an ocean model's output to the model parameters. Estimating this sensitivity is very time-consuming by brute force. Alternatively, adjoints have shown great promise in uncovering the sensitivity of the model to its parameters~\cite{mcnamara_fluid_nodate, errico1992sensitivity}. Yet, adjoints are very time consuming to develop manually and very involved to develop via automatic differentiation (AD) for some models. Because neural networks (NN) implemented in deep learning frameworks can be differentiated trivially, we have explored how to generate an accurate NN surrogate for an ocean model. 

We have considered the Simulating Ocean Mesoscale Activity (SOMA) test case for the MPAS-Ocean model and built neural network surrogates of the forward dynamics.
The contributions of this work can be summarized as follows. (1) We created a SOMA perturbed parameter ensemble dataset for deep learning model development and benchmarking; (2) We employed three different strategies to train neural network surrogates with large-scale distributed training, aiming to recreate the timestepping behavior of the forward (true) model. The trained models also showed consistent rollout performance for the midrange horizon; (3) we computed neural adjoints from the trained models and gained insight into the sensitivity to the varying Gent-McWilliams~(GM)~\cite{gent2011gent} parametrization.


\section{SOMA Test Case}
\label{sec:soma}
The Simulating Ocean Mesoscale Activity~(SOMA) experiment~\cite{soma} is a simulation within the Model for Prediction across Scales Ocean (MPAS-O)~\cite{RINGLER2013211,petersen2019,golaz2019}. SOMA simulates an eddying midlatitude ocean basin~(for details see Appendix~\ref{appx:soma}) with latitudes ranging
from 21.58 to 48.58N and longitudes ranging from 16.58W to 16.58E. The basin is circular and features curved coastlines
with a 150-km-wide, 100-m-deep continental shelf. SOMA is a more realistic version of typical double-gyre test cases, which are commonly used to assess idealized ocean model behavior.
We have chosen to estimate the isopycnal surface of the ocean at 32-km resolution. This diagnostic output is computed from five prognostic outputs that in turn are influenced by four model parameters. 

The original SOMA simulation runs for constant values of the scalar parameters that are being studied. For each parameter, a range was derived from the literature on reasonable values~(for details see Appendix~\ref{appx:soma}). 1000 samples within the range were drawn to form an ensemble. 
Each forward run involves using a parameter value from the sample, while using default values for the rest.
For each perturbed parameter run, the model is run for 2 years without recording any data. Then the model is run forward for one year, while recording the output at one-day intervals.

\section{Neural Network Surrogates}
\label{sec:surrogate}
\paragraph{Model Description}
On a high level, SOMA starts with the initial states, $x_0$, and the preset model parameters, $p$, and solves for the state variables at time $t$. With the initial value problem~(IVP),
$dx(t; p)/dt = f(x(t)); \quad x(0) = x_0$
we have the solution at time $t$,  
    $x(t) = x(0) + \int_{0}^{t} f(x(\tau; p))d \tau.$
The solution at the discrete time step $t+1$ can be expressed as 
\begin{equation}\label{eqn:one-step}
    x(t + 1) = x(t) + \int_{t}^{t+1}f(x(\tau; p))d \tau.
\end{equation}
The objective of building neural surrogates is to model the one-step solving process in~(\ref{eqn:one-step}).

Three types of neural networks commonly used in machine learning for climate liturature~\cite{nguyen2023climax, bonev2023spherical} were selected as surrogates for dynamics, namely Residual Network~(ResNet)~\cite{he2016deep}, U-Net~\cite{ronneberger2015u}, and Fourier Neural Operator~(FNO)~\cite{li2020fourier}. Resembling Euler method for solving ODEs, ResNets take advantage of identity mapping by skip connections, allowing deeper network training while mimicking the structure of (\ref{eqn:one-step}). U-Nets utilize skip connections in a different way, where identity mappings connect the first and last layers, the second and second-to-last layers, and so on. U-Nets are beneficial when network input and output have similar patterns. FNO, on the other hand, learns the solution operator to the IVP lying in infinite dimensional space via the composition of kernel integral operators in the Fourier domain. 

\paragraph{Training Strategy}
The network training for all models in this work followed the same strategy which included the choice of loss function, the application of the loss mask, the batch size, the number of epochs, and the learning rate. The loss function is the relative $L_p$ norm, shown as follows
$L_p = \Vert y - \hat{y} \Vert_p / \Vert y \Vert_p,$
where $y$ and $\hat{y}$ are the flattened true and predicted spatial-temporal varying state variables. Due to the shape of the domain of interest, a mask was applied to calculate the loss values during training, only considering the values within the domain. Training with data of three dimensions in space and multiple variables can be challenging. Therefore, we distributed data loading and network training using \texttt{PyTorch}~\cite{paszke2019pytorch} distributed training tools. Each model used 40 NVIDA A100 GPUs.

\paragraph{Evaluation Metrics}
We evaluated the performance of the one-step forward solving neural surrogates with two metrics, Coefficient of Determination~($R^2$) and symmetric Mean Aboslute Percentage Error~(sMAPE). $R^2$, defined as $R^2 = 1 - {\sum (y - \hat{y})^2}/{\sum(y - \Bar{y})^2}$ shows the variance ratio in the target variable~(output state variables) that can be explained by the learned models. A high value of $R^2$ is preferred. On the other hand, sMAPE, defined as $sMAPE = {\Vert y - \hat{y}\Vert_1}/{0.5(\Vert y + \hat{y} \Vert_1)}$, implies how much deviation, on average, the predicted values are from the ground truth. A lower value of sMAPE indicates better performance.

\section{Adjoint Computation}
\label{sec:adjoints}
To obtain the sensitivity of the neural surrogates to the parameters of the physical model, we calculated the Jacobian of the surrogate output by standard backpropagation in \texttt{PyTorch} at random locations. The output field is of shape $y \in \mathbb{R}^{n \times 100 \times 100 \times 60}$, where $n$ is the number of state variables. The input model parameters, which do not vary spatially, have the shape of $p \in \mathbb{R}^{m}$, where $m$ is the number of model parameters.  The actual full Jacobian $J = \frac{\partial y}{\partial p} \in \mathbb{R}^{n \times 100 \times 100 \times 60 \times m}$, which is large, memory intensive, and inefficient to calculate. Instead, we randomly pick a spatial location for all prognostic variables and calculate the Jacobian of the function at the specific location with respect to the inputs corresponding to the model parameters~(GM). As a result, the Jacobian was reduced to $J_{{loc.}} \in \mathbb{R}^{n \times m}$. These Jacobians were used to rank the sensitivity of output state variables to the GM, described in Section~\ref{sec:discussion}.




\section{Results and discussion}\label{sec:discussion}


We trained ResNet, U-Net, and FNO using the data with varied GM values obtained from SOMA forward runs. The dataset contained 100 forward runs with a different GM value for each, where each run contained a month of data. There were 80 runs randomly selected for training, 10 for validation and the rest 10 for testing purposes. We report the performance of the models on the testing set using the model checkpoints associated with the best performance on the validation set.
\begin{figure}
    \centering
    \begin{subfigure}{.32\linewidth}
        \centering
        \includegraphics[width=\linewidth]{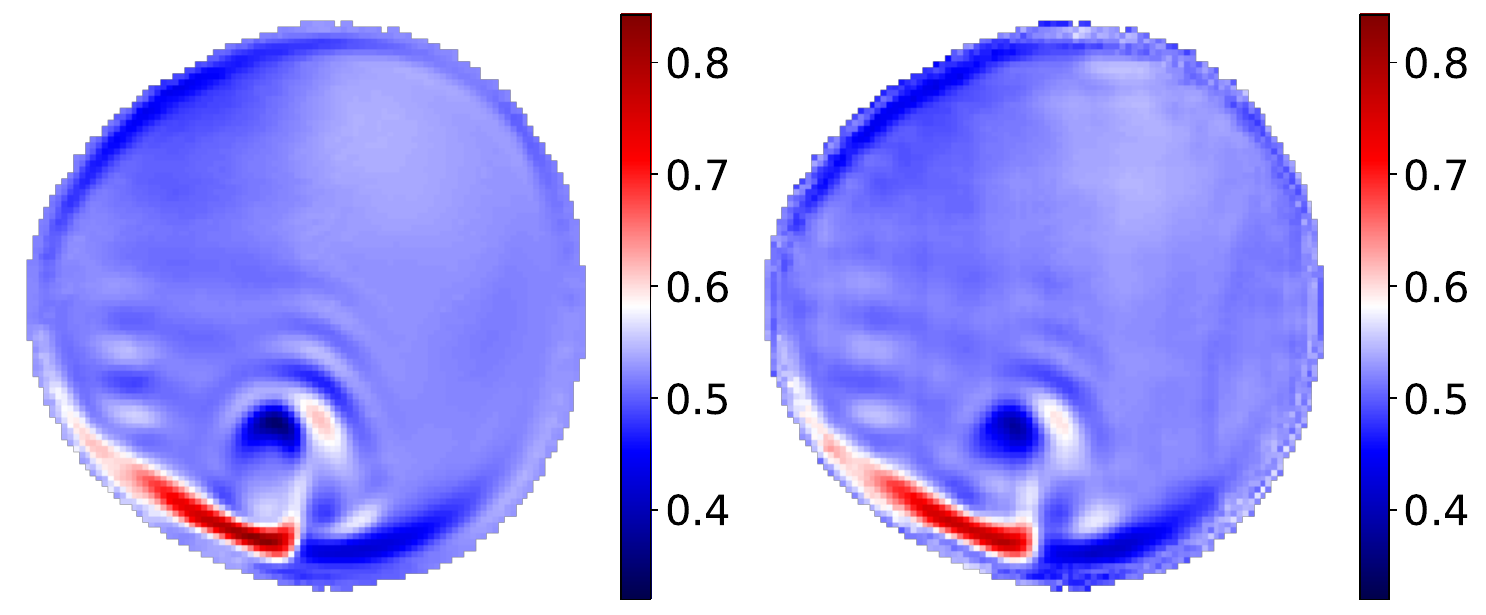}
        \caption{Zonal velocity.}
    \end{subfigure} 
    \begin{subfigure}{.32\linewidth}
        \centering
        \includegraphics[width=\linewidth]{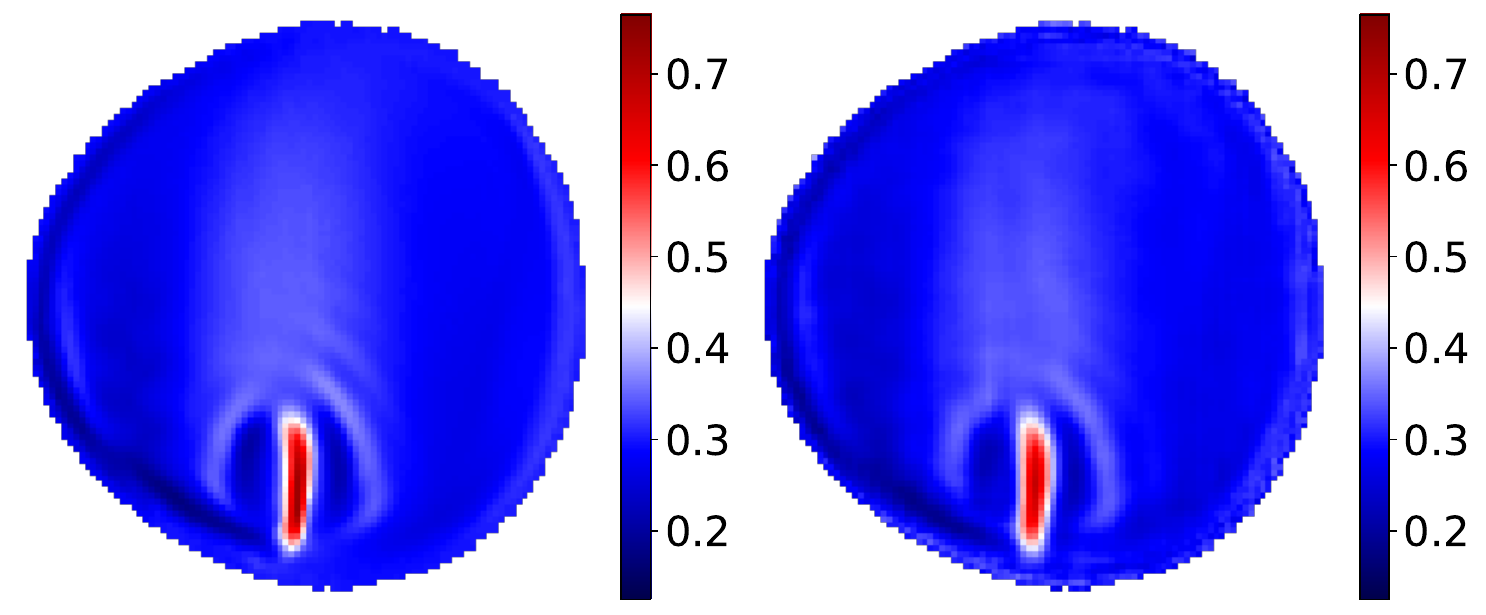}
        \caption{Meridional velocity.}
    \end{subfigure} 
    \begin{subfigure}{.32\linewidth}
        \centering
        \includegraphics[width=\linewidth]{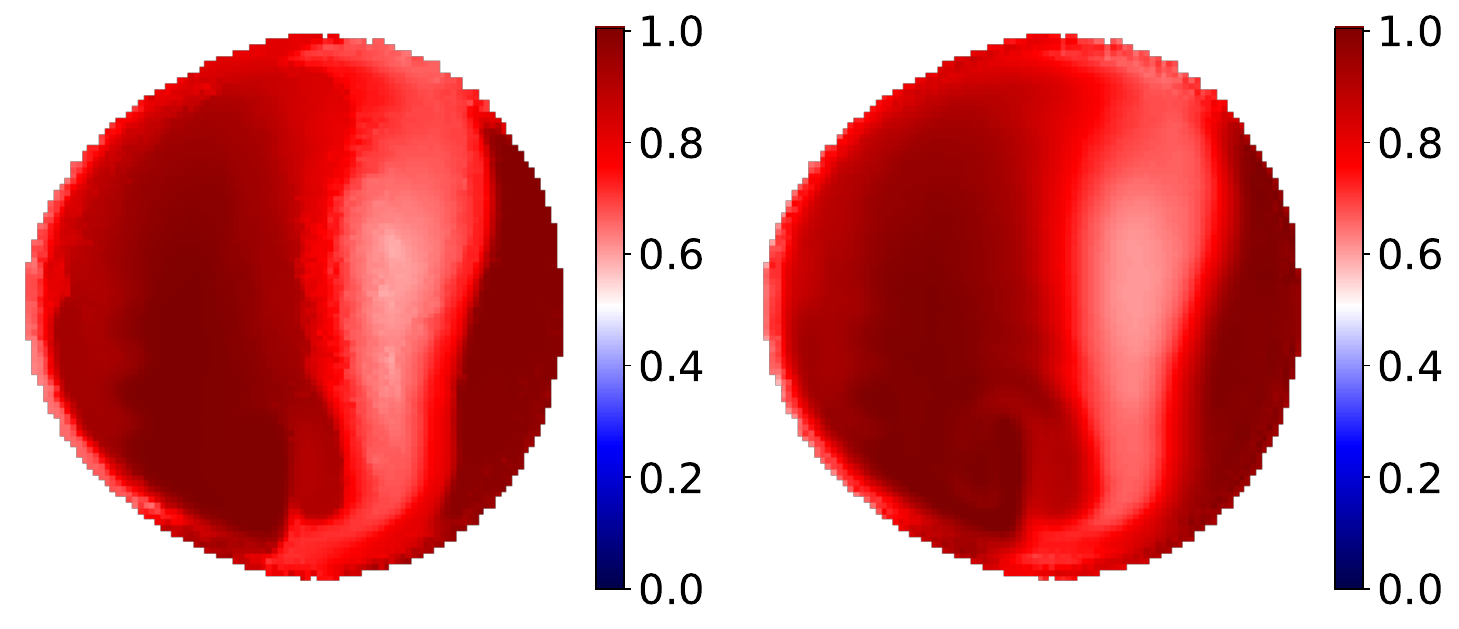}
        \caption{Temperature}
    \end{subfigure} 

    \begin{subfigure}{.32\linewidth}
        \centering
        \includegraphics[width=\linewidth]{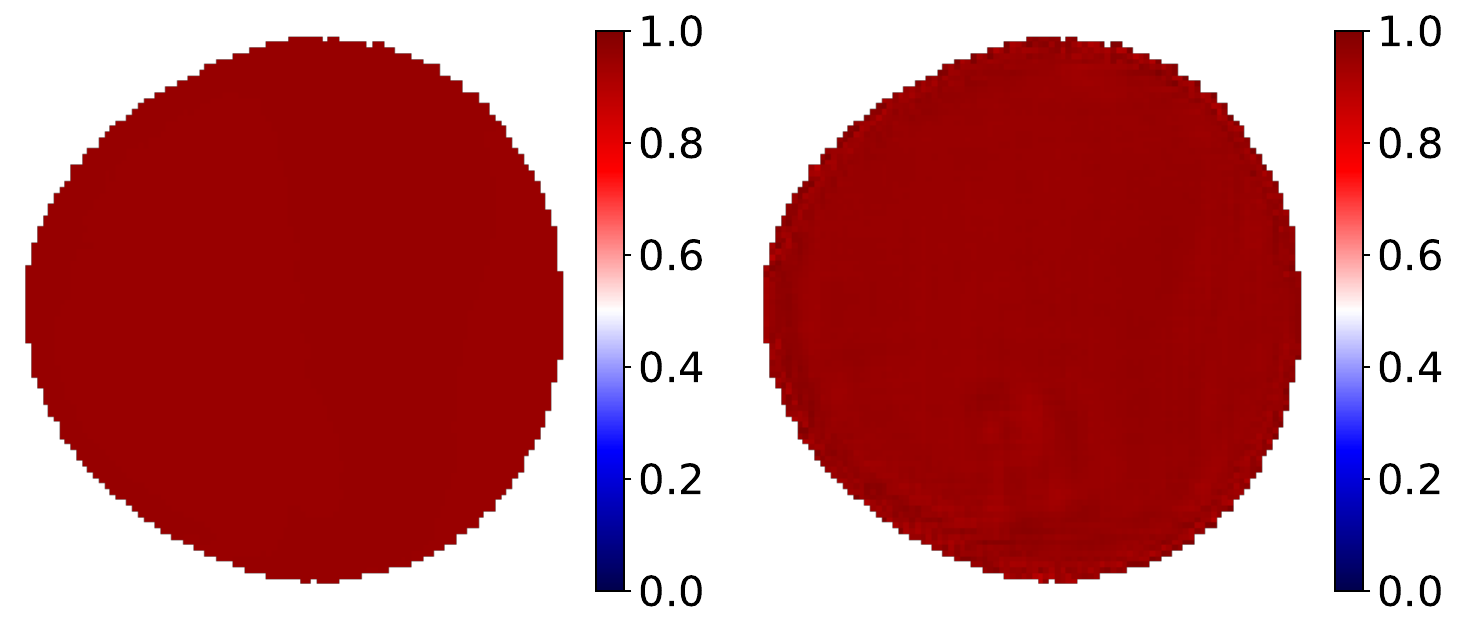}
        \caption{Salinity}
    \end{subfigure}
    \begin{subfigure}{.32\linewidth }
        \centering
        \includegraphics[width=\linewidth]{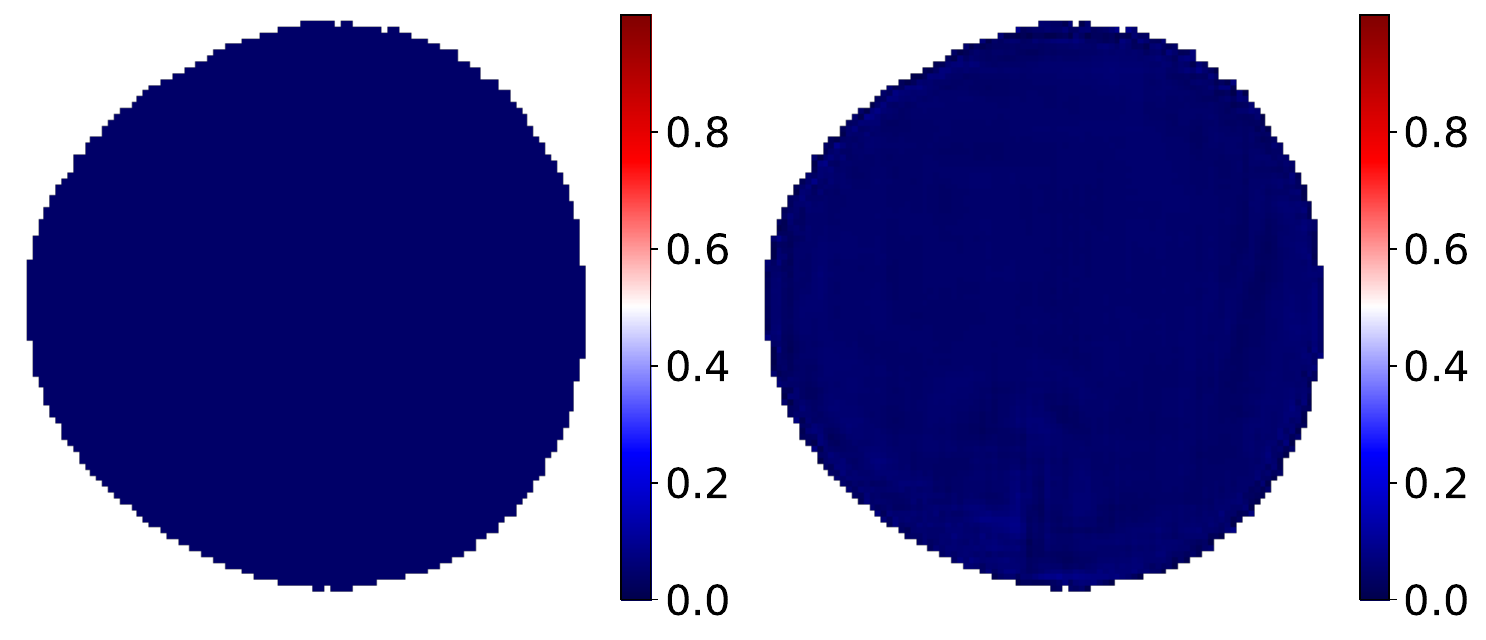}
        \caption{Layer thickness}
    \end{subfigure}
    \caption{The~(normalized) value of prognostic QOI at level 10 after 15 timesteps for the true model (left) and predicted by a UNet Neural Network surrogate (right). Here the NN model is used for a single timestep only.}\label{fig:onestep_res}
    \label{fig:UNet-true-pred}
\end{figure}

\begin{table}[]
    \centering
    \caption{Single Step Forward Prediction Results.} \label{tab:model_performance}
    \begin{subtable}{\textwidth}
    \centering
    \resizebox{\textwidth}{!}{
    \begin{tabular}{lccccccccc}
    \toprule
        ~ & $R^2$ & MAPE & relative & $R^2$ & MAPE & relative & $R^2$ & MAPE & relative\\
        ~ &  & (\%) & MSE (\%) &   & (\%) & MSE (\%) &   & (\%) & MSE (\%)\\
         \midrule
                            & \multicolumn{3}{c|}{UNet} & \multicolumn{3}{c|}{ResNet} & \multicolumn{3}{c|}{FNO}\\
        \midrule
        Layer thickness     & 0.9989 & 0.974 & 0.016 & 0.9989 & 3.319 & 2.259 & 0.9906 & 1.571 & 0.125 \\ 
        Salinity            & 0.9997 & 1.289 & 0.019 & 0.9901 & 9.140 & 5.649 & 0.9967 & 2.340 & 0.116\\ 
        Temperature         & 0.9968 & 3.440 & 0.081 & 0.9868 & 5.783 & 2.719 & 0.9968 & 1.658 & 0.044\\ 
        Zonal Vel.      & 0.9988 & 1.047 & 0.007 & 0.9914 & 1.857 & 6.137 & 0.9964 & 1.550 & 0.022 \\ 
        Meridional Vel. & 0.9969 & 4.444 & 0.002 & 0.9567 & 1.608 & 4.793 & 0.9893 & 0.893 & 0.008 \\ 
        \bottomrule
    \end{tabular}
    }
    \end{subtable}
\end{table}


\begin{figure}[tbhp]
    \centering
    \begin{minipage}[t]{0.6\textwidth}
        \vspace{0pt}
        \centering
        \begin{subfigure}[t]{.49\linewidth}
            \centering
            \includegraphics[width=\linewidth]{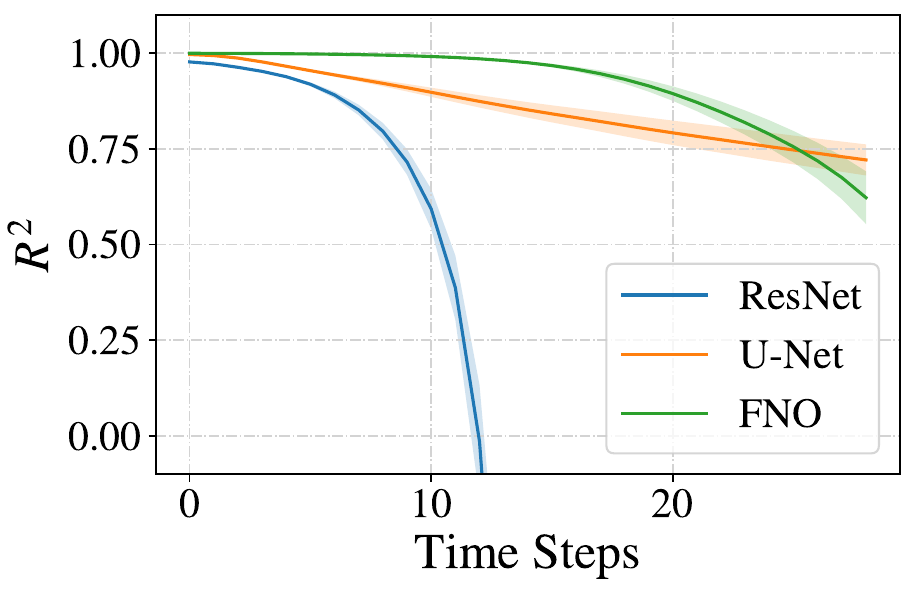}
            \caption{$R^2$: Temperature rollout.}
        \end{subfigure}
        \begin{subfigure}[t]{.49\linewidth}
            \centering
            \includegraphics[width=\linewidth]{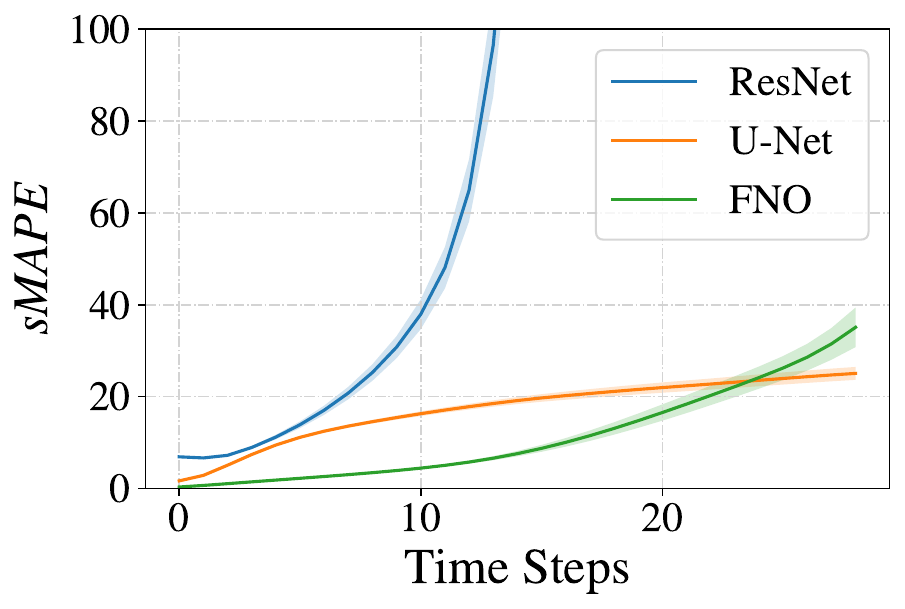}
            \caption{$sMAPE$: Temperature rollout.}
        \end{subfigure}
        \caption{Model rollout performance progression with horizon~(time steps). (a) and (b) show the performance scores of trained U-Net, ResNet, and FNO on temperature. FNO has the best performance among the three models, where the prediction errors in the first 10 steps are minor but quickly increase in the later steps.}
        \label{fig:rollout-main}
    \end{minipage}\hfill 
    \begin{minipage}[t]{0.37\textwidth}
        \vspace{0pt}
        \centering
        \includegraphics[width=\linewidth]{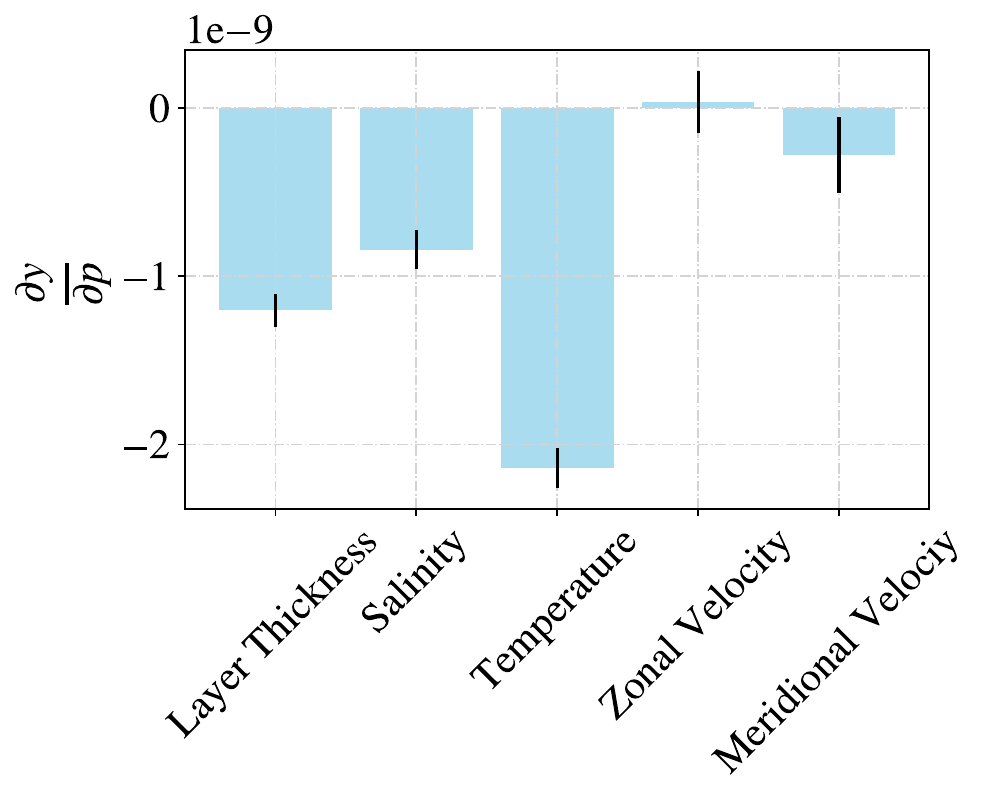}
        \caption{Average surrogate adjoint of prognostic variables to GM. Values were calculated using the test set at a fixed randomly picked location.}
        \label{fig:adjoint}
    \end{minipage}

\end{figure}

Table~\ref{tab:model_performance} shows the performance metrics of ResNet, U-Net, and FNO for each output prognostic variables~(layer thickness, salinity, temperature, zonal velocity, and meridional velocity). All three models accurately predict prognostic variables one step forward. Figure~\ref{fig:onestep_res} shows the true and predicted one-step forward values for the prognostic variables of the trained U-Net. The predicted fields closely resemble the ground truth, with a slight loss of details. To further investigate the performance of neural surrogates for a longer prediction horizon, we applied the trained models in an autoregressive way to produce rollouts for the entire month.  

Figure~\ref{fig:rollout-main} presents the rollout performance of trained U-Net, ResNet, and FNO on meridional velocity and temperature in the test set. Among the trained models, FNO outperforms the other two in terms of longer-horizon rollout. FNO results in a high accuracy for the first 10 steps and quickly degrades as the error accumulates rapidly. In contrast, U-Net and ResNet perform worse than FNO. In particular, the error introduced by ResNet increases rapidly from the beginning, making it the worst performing model in rollout. On the basis of this observation, we focus the subsequent work on FNO and its variants.

As a first step to understand the model adjoint sensitivity to the model parameters, we computed the surrogate adjoint from the trained FNO. Figure~\ref{fig:adjoint} shows the average neural surrogate output derivatives with respect to the parameter~(GM). The results suggest that GM has the most significant impact on temperature among the prognostic variables, followed by layer thickness, salinity, and meridional velocity. Meanwhile, the zonal velocity was least affected. Although the neural surrogates emulated the physical simulation in one-step forward solving, due to the enormous search space of the trainable weights and fixed data resolution, it is possible that the adjoint from the neural surrogates did not match with them of the physical model. Therefore, we propose to perform the dot-product test, described in Appendix~\ref{app:dot-test}, to verify it in future work.

\section{Conclusion}
\label{sec:conclusion}
The neural surrogates we developed, based on perturbed parameters from SOMA runs, accurately predict the prognostic variables in one-step forward compared to the original SOMA model. Additionally, we have been successful in computing the neural adjoints from the trained models and obtained initial insights on the senstivities. Our future work includes verifying neural adjoints against the approximated true adjoints via the dot-product test, improving adjoint-aware training by incorperating known physics, and investigating the feasibility of applying our methodology to the MPAS-O code, specifically in a configuration simulating the AMOC. We also intend to implement our approach in conjunction with a SOMA-like MITgcm configuration.

\begin{ack}
We gratefully acknowledge the computing resources provided on Bebop, a high-performance computing  cluster operated by LCRC at Argonne National Laboratory.
This research used resources of the NERSC, a U.S. Department of Energy Office of Science User Facility located at LBNL.
Material based upon work supported by the US DoE, Office of Science, Office of Advanced Scientific Computing Research and Office of BER, Scientific Discovery through Advanced Computing (SciDAC) program, under Contract DE-AC02-06CH11357. We are grateful to the Sustainable Horizons Institute's Sustainable Research Pathways workforce development program.
\end{ack}

\bibliographystyle{unsrt}
\bibliography{refs}
\vfill

\pagebreak
\appendix
\section{Ongoing and future work}\label{app:dot-test}
\textit{Dot-product Test.}
The Jacobian of the neural surrogate at the location has the form $J = \frac{\partial y}{\partial h_N} \frac{\partial h_N}{\partial h_{N-1}} \dots \frac{\partial h_0}{\partial p} \in \mathbb{R}^{n \times m}$, where $h$s are the outputs of the hidden layers. Surrogate adjoints are easily accessible by differentiation of trained neural networks. However, they may be drastically different from the actual adjoints from the physical forward model due to the enormous search space of the trainable weights and finite resolution of the data. To investigate the correctness of surrogate adjoints, we perform the dot-product test with two randomly generated vectors, using the approximated Jacobian via finite difference of the physical model and surrogate Jacobain computed via reverse-mode automatic differentiation of the trained neural networks. In particular, with random vectors $v \in \mathbb{R}^{p}$ and $w \in \mathbb{R}^{m}$, we test whether the equality sign holds in~(\ref{eqn:dotp-test}). 
\begin{equation}\label{eqn:dotp-test}
    \begin{aligned}
        &w^{\top}\cdot(\underbrace{\frac{\partial \mathcal{M}}{\partial p}v}_{\text{JVP}}) \stackrel{?}{=} (\underbrace{\underbrace{w^{\top}\frac{\partial y}{\partial h_N}} \frac{\partial h_N}{\partial h_{N-1}} \dots \frac{\partial h_0}{\partial p}}_{\text{VJP}})\cdot v, \\
        &\frac{\partial \mathcal{M}}{\partial p} \approx \frac{\mathcal{M}(x;p + \frac{1}{2}\Delta p) - \mathcal{M}(x; p - \frac{1}{2}\Delta p)}{\Delta p}.\\
    \end{aligned}
\end{equation}
Here, $\mathcal{M}$ is the nondifferentiable physical model used for running simulations.

\textit{Incorporating known physics.} The governing equations of SOMA or MPAS-O are well established. Utilizing known physics in the training of neural surrogates helps improve accuracy, reduce the requirement of large data size, and regularize learning for better generalization~\cite{li_physics-informed_2022, karniadakis2021physics, raissi_physics_2017, liu2022predicting}. We plan to incorporate the forms of physical inductive biases presented in~\cite{li_physics-informed_2022, liu2022predicting} into our existing models and hypothesize that training with known physics improves the neural surrogate adjoint matching.

\section{Rollout performance of all prognositic variables}
\begin{figure}[tbhp]
    \centering
    \begin{subfigure}{.4\linewidth}
        \centering
        \includegraphics[width=\linewidth]{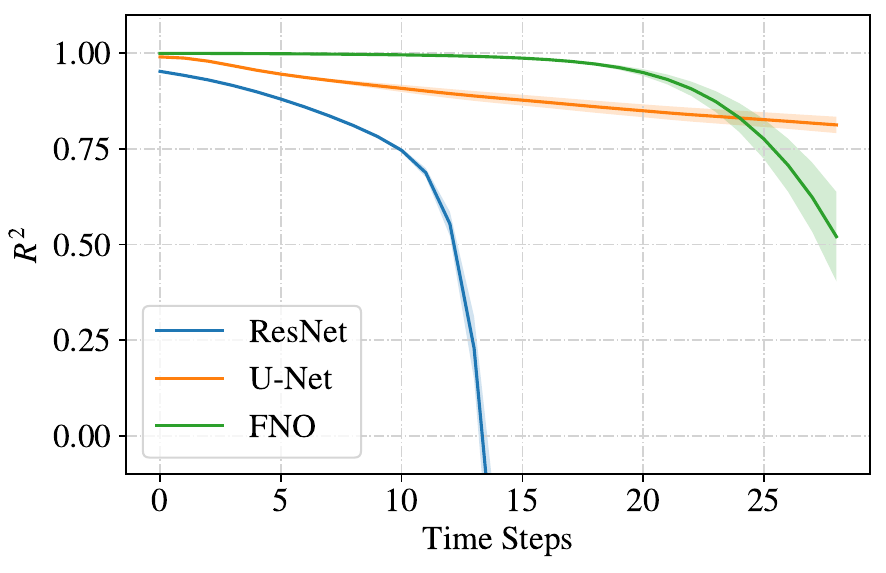}
        \caption{$R^2$: Layer Thickness rollout}        
    \end{subfigure}
    \begin{subfigure}{.4\linewidth}
        \centering
        \includegraphics[width=\linewidth]{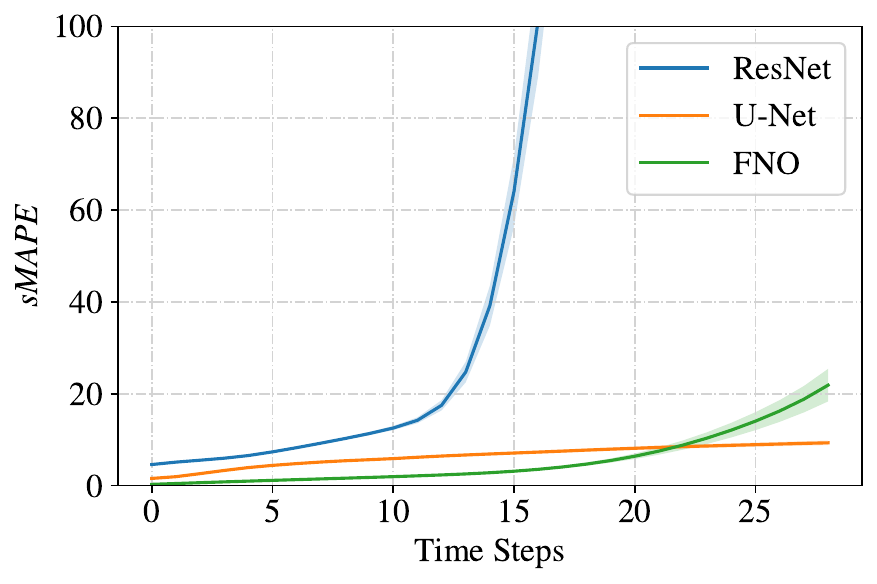}
        \caption{$sMAPE$: Layer Thickness rollout}        
    \end{subfigure}

    \begin{subfigure}{.4\linewidth}
        \centering
        \includegraphics[width=\linewidth]{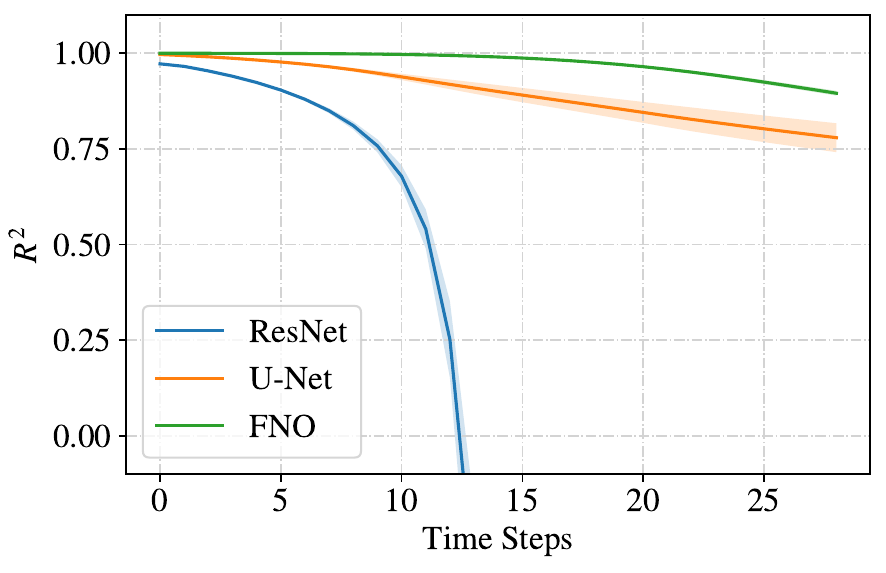}
        \caption{$R^2$: Salinity rollout}        
    \end{subfigure}
    \begin{subfigure}{.4\linewidth}
        \centering
        \includegraphics[width=\linewidth]{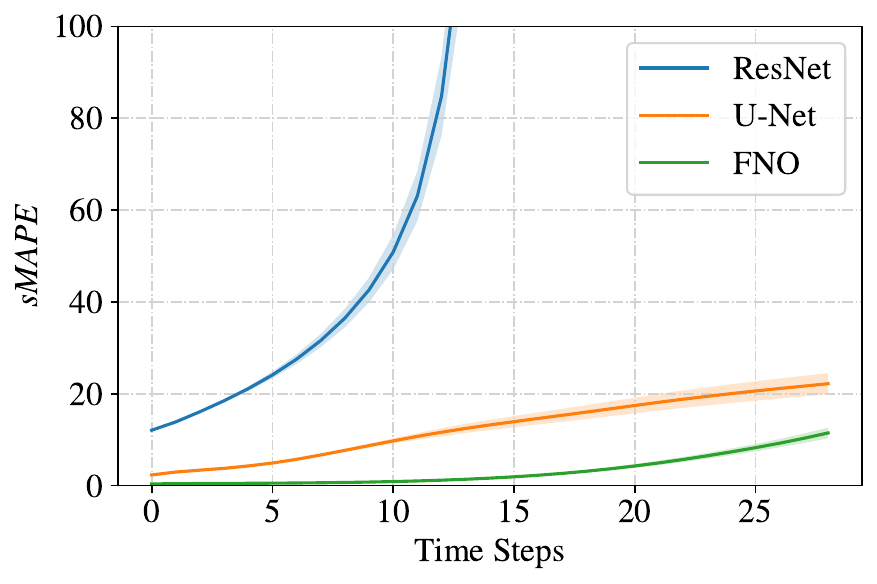}
        \caption{$sMAPE$: Salinity rollout}        
    \end{subfigure}
    
    \begin{subfigure}{.4\linewidth}
        \centering
        \includegraphics[width=\linewidth]{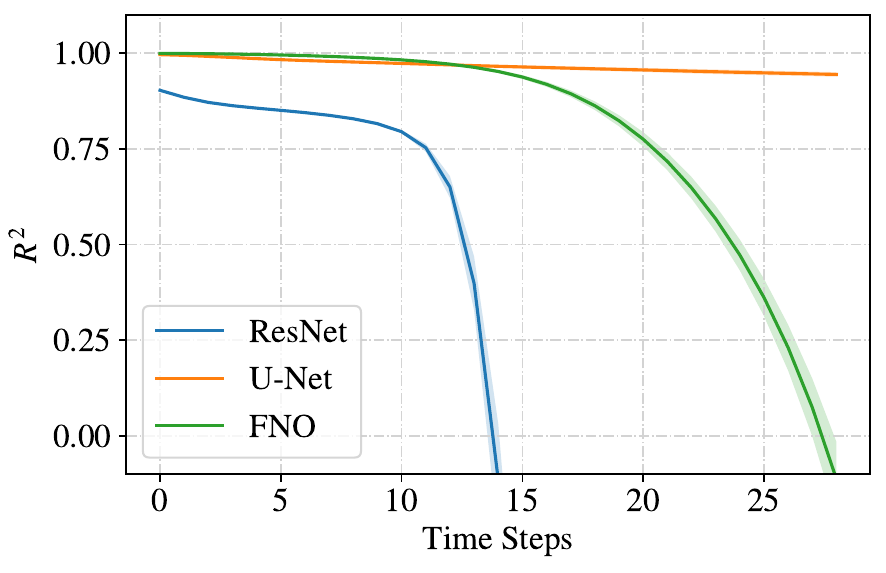}
        \caption{$R^2$: Zonal Velocity rollout}        
    \end{subfigure}
    \begin{subfigure}{.4\linewidth}
        \centering
        \includegraphics[width=\linewidth]{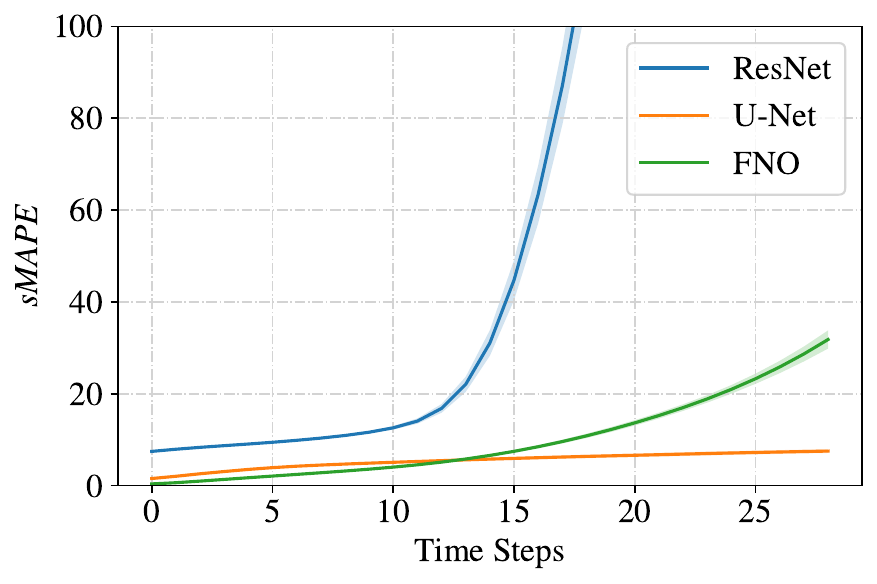}
        \caption{$sMAPE$: Zonal Velocity}        
    \end{subfigure}
    
    \begin{subfigure}{.4\linewidth}
        \centering
        \includegraphics[width=\linewidth]{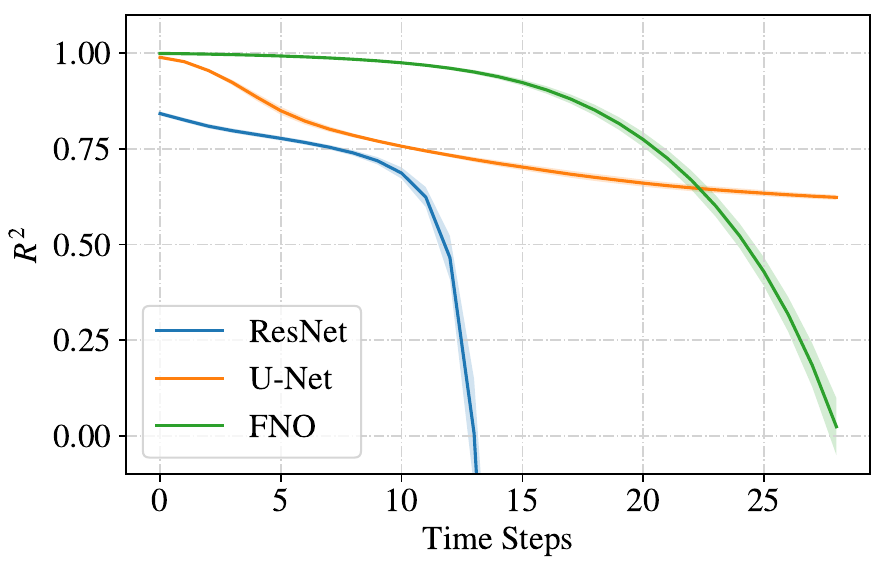}
        \caption{$R^2$: Meridional Velocity rollout.}        
    \end{subfigure}
    \begin{subfigure}{.4\linewidth}
        \centering
        \includegraphics[width=\linewidth]{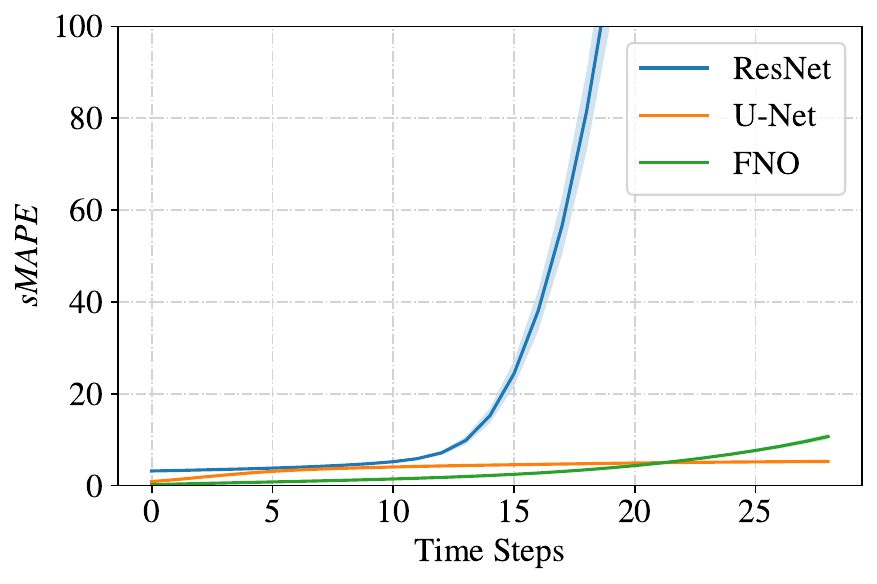}
        \caption{$sMAPE$: Meridional Velocity rollout.}        
    \end{subfigure}

    \begin{subfigure}{.4\linewidth}
        \centering
        \includegraphics[width=\linewidth]{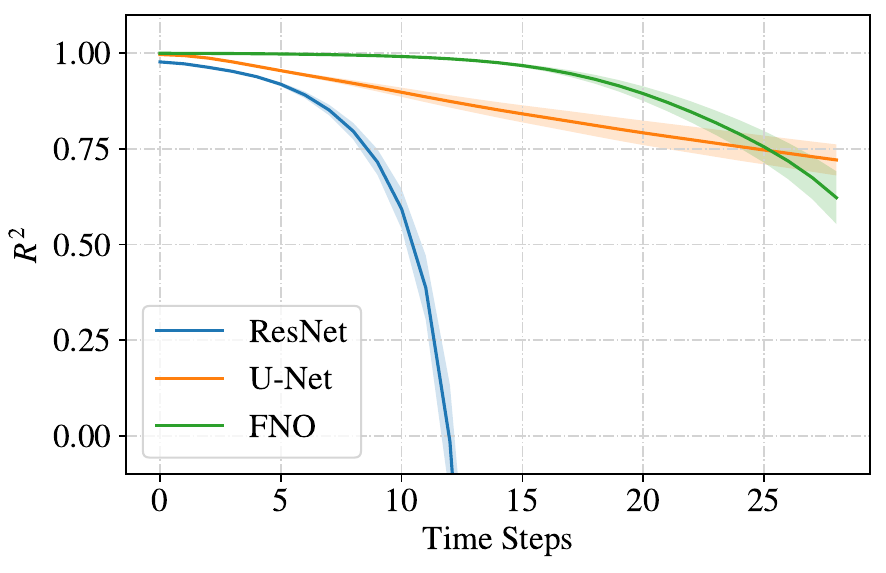}
        \caption{$R^2$: Temperature rollout.}        
    \end{subfigure}
    \begin{subfigure}{.4\linewidth}
        \centering
        \includegraphics[width=\linewidth]{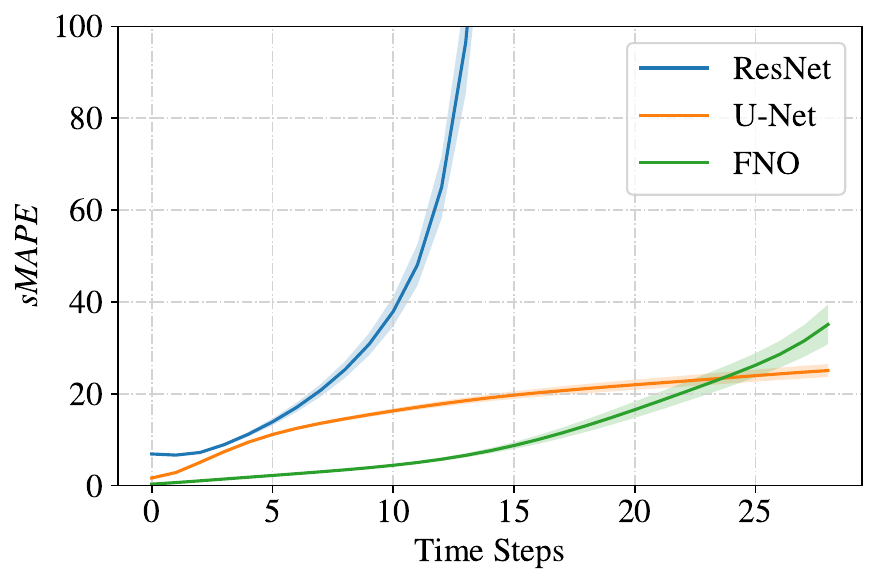}
        \caption{$sMAPE$: Temperature rollout.}        
    \end{subfigure}
    \label{fig:rollout}
    \caption{Rollout performance progression with horizon~(time steps) for the five prognostic variables} \label{fig:rollout}
\end{figure}

\pagebreak

\section{SOMA configuration}\label{appx:soma}


The SOMA configuration is designed to investigate equilibrium mesoscale activity in a setting similar to how ocean climate models are deployed. SOMA is used to represent an idealized,
eddying, midlatitude, double-gyre system. 
It simulates an eddying, midlatitude ocean basin with latitudes ranging
from 21.58 to 48.58N and longitudes ranging from 16.58W to 16.58E. The basin is circular and features curved coastlines
with a 150-km-wide, 100-m-deep continental shelf. SOMA can be run at four different resolutions, where a smaller resolution is more granular: 4km, 8km, 16km, and 32km.

We have chosen to estimate the isopycnal surface of the ocean. This prognostic output is computed from five diagnostic outputs which in turn are influenced by four model parameters. 

The original SOMA simulation runs for constant values of the scalar parameters that are being studied. For each parameter, a range was derived from literature on reasonable values. 1000 samples within the range were drawn to form an ensemble. 
Table~\ref{tab:parameterrange} lists the parameters as well as the maximum and minimum values for uniform sampling. Each forward run involves using a parameter value from the sample while using default values for the rest.
For each perturbed parameter run, the model is run for 2-years without recording any data. Then the model is run forward for 1-year while recording the output at 1-day intervals. Not all combinations of parameters resulted in convergence. 

\begin{table}[h]
    \centering
    \caption{Range of Perturbed Parameter Values}
    \begin{tabular}{ccc}

    \toprule
    \textbf{Parameter} & \textbf{Minimum} & \textbf{Maximum}\\
    \midrule
      GM\_constant\_kappa   & 200.0 & 2000.0 \\

      Redi\_constant\_kappa   & 0.0 & 3000.0\\

      cvmix\_background\_diff & 0.0 & 1e-4\\ 

      implicit\_bottom\_drag & 1e-4 & 1e-2\\
    \bottomrule
    \end{tabular}
    
    \label{tab:parameterrange}
\end{table}

\begin{figure}[h]
    \centering
    \includegraphics[width=.8\linewidth]{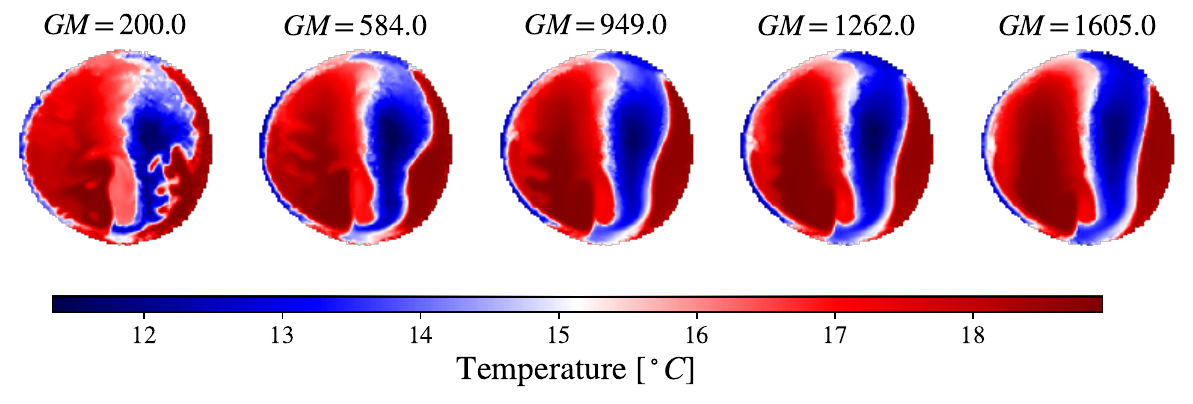}
    \caption{Variation shown in temperature with different values of Gent-McWilliams parametrization. The temperature at vertical level 10 at end of simulations of the same initial condition is shown. }\label{fig:gm_parametrization}
\end{figure}

At the 32km resolution, there are 8,521 hexagonal cells on the grid, each with 60 vertical layers, resulting in over 15 million data entries for each spatially and temporally varying output variable in the data set for the year. Finally, the data generated for mesh grid was converted to a standard latitude and longitude grid through spatial interpolation. 
The obtained raw data was converted from a mesh grid to a standard latitude and longitude grid through spatial interpolation.
Examining the data shows marked variation in output for different parameter values. Figure~\ref{fig:gm_parametrization} shows the variation shows in ocean temperature for different values of the Gent-McWilliams parametrization.
\begin{figure}[h]
    \centering
    \includegraphics[width=.8\linewidth]{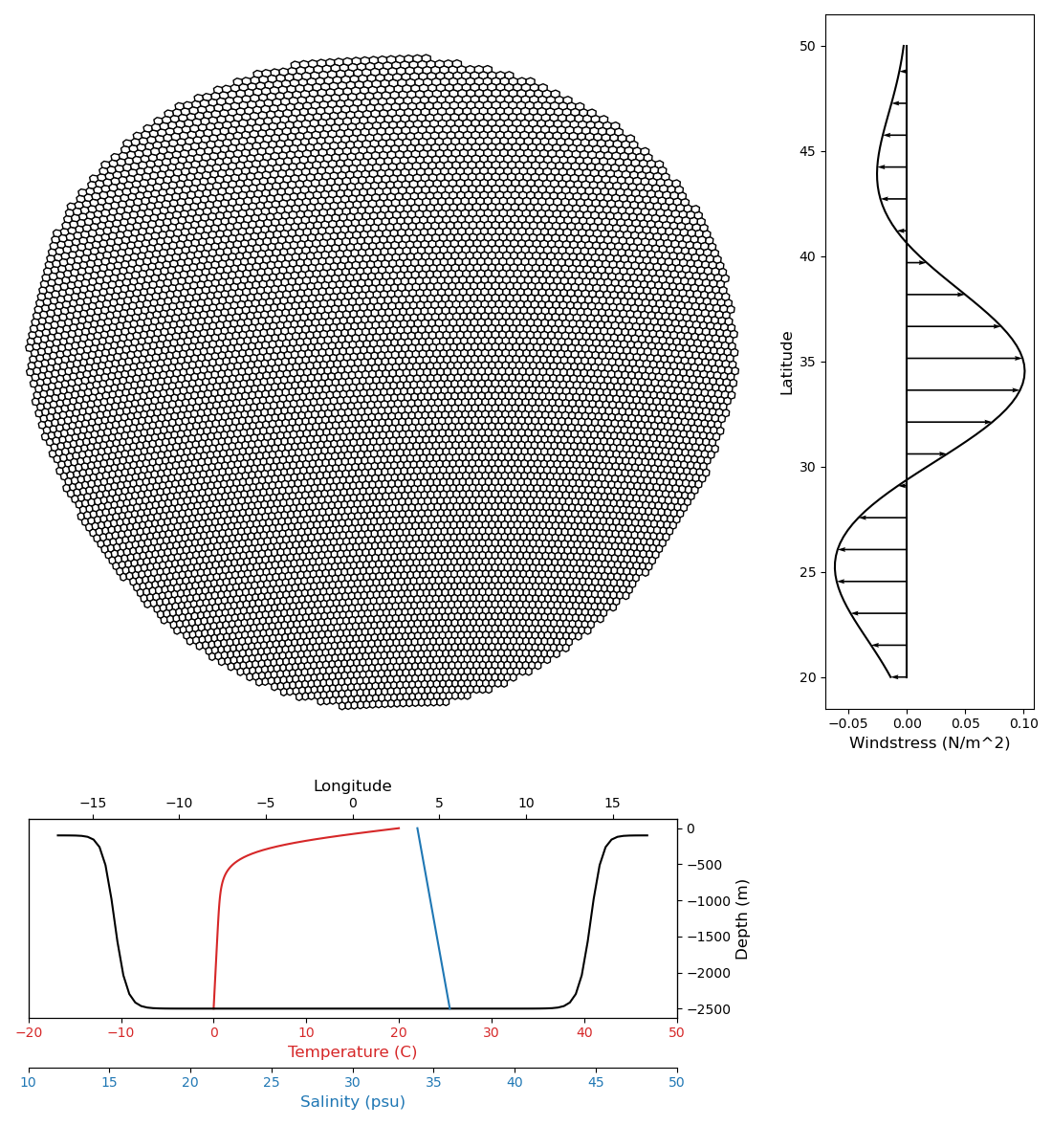}
    \caption{The SOMA domain is shown with the 32km mesh. Below is the depth profile of the basin along with the horizontally constant initial temperature and salinity profiles. To the right is the longitudinally-constant imposed wind stress forcing.}        
\end{figure}

\end{document}